\documentclass[graybox]{svmult}
\usepackage[utf8]{inputenc}
\usepackage[T1]{fontenc}
\usepackage{mathptmx}       
\usepackage{helvet}         
\usepackage{courier}        
\usepackage{makeidx}         
\usepackage{graphicx}        
\graphicspath{{./Figures/}}        %
\usepackage{multicol}        
\usepackage[bottom]{footmisc}
\usepackage{natbib}
\usepackage{hyperref}
\usepackage{amsmath}
\RequirePackage{newunicodechar}
\newunicodechar{°}{\ensuremath{^{\circ}}}
\author{Marko Laine}
\date{\today}
\title{Introduction to Dynamic Linear Models for Time Series Analysis}
\hypersetup{
 pdfauthor={Marko Laine},
 pdftitle={Introduction to Dynamic Linear Models for Time Series Analysis},
 pdfkeywords={DLM},
 pdfsubject={Dynamic linear model introduction},
 pdfcreator={Emacs 26.2 (Org mode 9.1.9)}, 
 pdflang={English}}
\begin{document}

\institute{Marko Laine \at Finnish Meteorological Institute, Helsinki, Finland, \email{marko.laine@fmi.fi}}

\maketitle

\abstract{
Dynamic linear models (DLM) offer a very generic framework to analyse time series data. Many classical time series models can be formulated as DLMs, including ARMA models and standard multiple linear regression models. The models can be seen as general regression models where the coefficients can vary in time. In addition, they allow for a state space representation and a formulation as hierarchical statistical models, which in turn is the key for efficient estimation by Kalman formulas and by Markov chain Monte Carlo (MCMC) methods. A dynamic linear model can handle non-stationary processes, missing values and non-uniform sampling as well as observations with varying accuracies. This chapter gives an introduction to DLM and shows how to build various useful models for analysing trends and other sources of variability in geodetic time series.
}

\newcommand{\nobs}{n}
\newcommand{\abs}[1]{\left\lvert#1\right\rvert}
\newcommand{\state}{\vec{x}}
\newcommand{\param}{\theta}
\newcommand{\widebar}{\overline}
\DeclareRobustCommand*{\vec}[1]{\ensuremath{%
\mathchoice{\mbox{\boldmath$\displaystyle#1$}}
           {\mbox{\boldmath$\textstyle#1$}}
           {\mbox{\boldmath$\scriptstyle#1$}}
           {\mbox{\boldmath$\scriptscriptstyle#1$}}}}
\DeclareRobustCommand*{\mat}[1]{\ensuremath{%
\mathchoice{\mbox{\boldmath$\mathrm{\displaystyle#1}$}}
           {\mbox{\boldmath$\mathrm{\textstyle#1}$}}
           {\mbox{\boldmath$\mathrm{\scriptstyle#1}$}}
           {\mbox{\boldmath$\mathrm{\scriptscriptstyle#1}$}}}}
\setcounter{chapter}{3}

\section{Introduction to dynamic linear models}
\label{sec:org13245d7}
Statistical analysis of time series data is usually faced with the fact that we have only one realization of a process whose properties might not be fully understood. We need to assume that some distributional properties of the process that generate the observations do not change with time. In linear trend analysis, for example, we assume that there is an underlying change in the background mean that stays approximately constant over time. Dynamic regression avoids this by explicitly allowing temporal variability in the regression coefficients and by letting some of the system properties to change in time. Furthermore, the use of unobservable state variables allows direct modelling of the processes that are driving the observed variability, such as seasonal variation or external forcing, and we can explicitly allow some modelling error.

Dynamic regression can  be formulated in very general terms by using a state space representation of the observations and the hidden state of the system. With sequential definition of the processes, having conditional dependence only on the previous time step, the classical recursive Kalman filter algorithms can be used to estimate the model states given the observations. When the operators involved in the definition of the system are linear we have so called dynamic linear model (DLM).

A basic model for time series in geodetic or more general environmental applications consists of four elements: a slowly varying background level, a seasonal component, external forcing from known processes modelled by proxy variables, and stochastic noise. The noise component might contain an autoregressive structure to account for temporally correlated model residuals. As we see, the basic components have some physical justification and we might be interested in their contribution to the overall variability and their temporal changes. These components are hidden in the sense that we do not observe them directly and each individual component is masked by various other sources of variability in the observations.

Below, we briefly describe the use of dynamic linear models in time series analysis. The examples deal with univariate time series, i.e.\  the observation at a singe time instance is a scalar, but the framework and the computer code can handle multivariate data, too. All the model equations are written in way that support multivariate observations. In the presented applications we are mostly interested in extracting the components related to the trends and using these to infer about their magnitude and the uncertainties involved. However, these models might not be so good for produce predictions about the behaviour of the system in the future, although understanding the system is a first step to be able to make predictions.

The use of DLMs in time series analysis is well documented in statistical literature, but they might go by different terminology and notation. In \cite{harvey91} they are called structural time series, \cite{durbin2012} uses the state space approach, and the acronym DLM is used in \cite{petris2009}.

\section{State space description}
\label{sec:orgc25efeb}
The state space description offers a unified formulation for the analysis of dynamic regression models. The same formulation is used extensively in signal processing and geophysical data assimilation studies, for example. A general dynamic linear model with an observation equation and a model equation is
\begin{eqnarray}
\label{dlmDef1}
  y_{t} &=& \mat{H}_{t}x_{t}+\epsilon_{t}, \quad  \epsilon_{t}\sim
  N(0,\mat{R}_{t}), \\ \label{dlmDef2}
  x_{t} &=& \mat{M}_{t}x_{t-1}+E_{t},\quad  E_{t}\sim N(0,\mat{Q}_{t}).
\end{eqnarray}

Above \(y_{t}\) is a vector of length \(k\) of observations at time \(t\), with \(t=1,\dots,n\). Vector \(x_{t}\) of length \(m\) contains the unobserved states of the system that evolve in time according to a linear \emph{system operator} \(\mat{M}_{t}\) (a \(m\times m\) matrix). In time series settings \(x_t\) will have elements corresponding to various components of the time series process, like trend, seasonality, etc. We observe a linear combination of the states with noise \(\epsilon_{t}\), and matrix \(\mat{H}_{t}\) (\(m\times k\)) is the \emph{observation operator} that transforms the model states into observations. Both observations and the system states can have additive Gaussian errors with covariance matrices \(\mat{R}_{t}\) (\(k\times k\)) and \(\mat{Q}_{t}\) (\(m\times m\)), respectively. In univariate time series analysis we will have \(k=1\). With multivariate data, the system matrices \(\mat{M}_{t}\), \(\mat{H}_{t}\), \(\mat{R}_{t}\) and \(\mat{Q}_{t}\) can be used to define correlations between the observed components.

This formulation is quite general and flexible as it allows handling of many time series analysis problems in a single framework. Moreover, a unified computational tool can be used, i.e.\  a single DLM computer code can be used for various purposes. Below we give examples of different analyses. As we are dealing with linear models, we assume that the operators \(\mat{M}_{t}\) and \(\mat{H}_{t}\) are linear. However, they can change with the time index \(t\) and we will drop the time index in the cases where the matrices are assumed static in time. The state space framework can be extended to non-linear model and non-Gaussian errors, and to spatial-temporal analyses as well, see, e.g., \cite{cressie11,sarkka13}. However, as can be seen in the following example, already the dynamic linear Gaussian formulation provides a large class of models for time series trend analyses.

\subsection{Example: spline smoothing}
\label{sec:org0f86055}
A simple local level and local trend model can be used as a basis for many trend related studies. Consider a mean level process \(\mu_{t}\) which is changing smoothly in time and which we observe with additive Gaussian noise. We assume that the change in the mean, \(\mu_{t+1}-\mu_{t}\), is controlled by a trend process \(\alpha_{t}\) and the temporal change in these processes is assumed to be Gaussian with given variances \(\sigma^2_\mathrm{level}\) and \(\sigma^2_\mathrm{trend}\). This can be written as
\begin{align}
  y_t &= \mu_{t} + \epsilon_\mathrm{obs}, &\epsilon_\mathrm{obs}\sim N(0,\sigma^2_\mathrm{obs}),  &\text{ observations,}\\
  \mu_{t} &= \mu_{t-1} + \alpha_{t-1} + \epsilon_\mathrm{level},
  & \epsilon_\mathrm{level}\sim N(0,\sigma^2_\mathrm{level}), &\text{ local level,}\\
  \alpha_{t} &=\alpha_{t-1} +\epsilon_\mathrm{trend},
 &\epsilon_\mathrm{trend}\sim N(0,\sigma^2_\mathrm{trend}),
 &\text{ local trend},
\end{align}
which in state space representation transfers simply into
\begin{equation}
\quad x_t = \begin{bmatrix} \mu_{t}&\alpha_{t}\end{bmatrix}^T,\quad
\mat{H}= \begin{bmatrix}1 & 0 \\\end{bmatrix},\quad
\mat{M}=\begin{bmatrix}
1 & 1 \\
0 & 1 \\
\end{bmatrix},
\end{equation}
\begin{equation}
\mat{Q}=\begin{bmatrix}
 \sigma^2_\mathrm{level} & 0\\
 0 &\sigma^2_\mathrm{trend} \\
\end{bmatrix},\quad\text{and}\quad
\mat{R}= \mathrm{diag}\left(\begin{bmatrix}\sigma^2_\mathrm{obs} & \dots & \sigma^2_\mathrm{obs}\end{bmatrix}\right),
\end{equation}
with three parameters for the error variances
\begin{equation}
\theta =
  \begin{bmatrix}
     \sigma^2_\mathrm{obs}&\sigma^2_\mathrm{level} &\sigma^2_\mathrm{trend}
  \end{bmatrix}^T.
\end{equation}
We have dropped the time index \(t\) from those elements that do not depend on time.

It is interesting to note, that if we set \(\sigma^2_\mathrm{level} = 0\), we have a second difference process for \(\mu_{t}\) as
\begin{equation}
\Delta^{2}\mu_{t} = \mu_{t-1} - 2\mu_{t} + \mu_{t+1} = \Delta\alpha_{t} = \epsilon_\mathrm{trend},
\end{equation}
and it can be shown \citep{durbin2012} that this is equivalent to cubic spline smoothing  with smoothing parameter \(\lambda = \sigma^2_\mathrm{trend}/\sigma^2_\mathrm{obs} > 0\).

\begin{figure}[htbp]
\centering
\includegraphics[width=.9\linewidth]{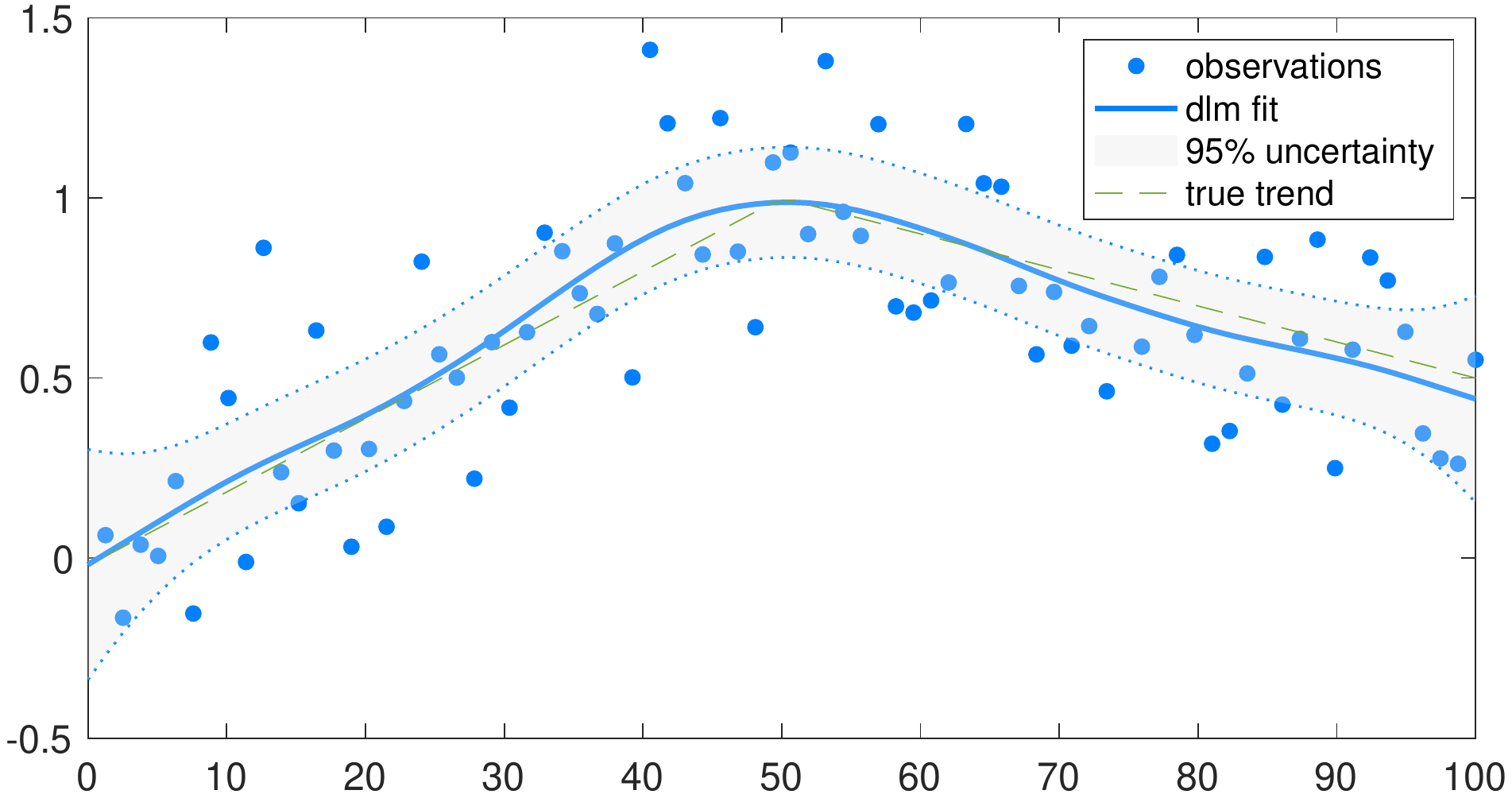}
\caption{\label{fig:org351a4f1}
DLM smoother fit to synthetic data set using a local trend model. In this example \(\sigma_{\mathrm{obs}}=0.3\), \(\sigma_{\mathrm{level}}=0.0\), and \(\sigma_{\mathrm{trend}}=0.01\), with time interval equal to one unit.}
\end{figure}

Figure \ref{fig:org351a4f1} shows simulated observations with a true piecewise trend and the fitted mean process \(\mu_{t}\), \(t=1,\dots,n\) together with its 95\% uncertainty limits. In this example, the observation uncertainty standard deviation (\(\sigma_{\mathrm{obs}}=0.3\)) as well as the level and trend variability standard deviations (\(\sigma_{\mathrm{level}}=0.0\), \(\sigma_{\mathrm{trend}}=0.01\)) are assumed to be known. In the later examples these values are estimated from the data. 

\section{DLM as hierarchical statistical model}
\label{sec:orgf892545}
The DLM formulation can be seen as a special case of a general hierarchical statistical model with three levels: data, process and parameters (see e.g. \cite{cressie11}), with corresponding conditional statistical distributions. First, the observation uncertainty \(p(y_{t}\vert x_{t},\theta)\) described by the observation equation and forming the statistical likelihood function, second, the process uncertainty of the unknown states \(x_{t}\) and their evolution given by the process equations as \(p(x_{t}\vert \theta)\) or \(p(x_{t}\vert x_{t-1},\theta)\), and third, the unconditional prior uncertainty for the model parameters \(p(\theta)\). This formulation allows both an efficient description of the system and computational tools to estimate the components. It also combines different statistical approaches, as we can have full prior probabilities for the unknowns (the Bayesian approach), estimate them by maximum likelihood and plug them back (frequentistic approach), or even fix the model parameters by expert knowledge (a non-statistical approach). By the Bayes formula, we can write the state and parameter posterior distributions as a product of the conditional distributions
\begin{equation}\label{equ:bayes}
p(x_{t},\theta \vert y_{t}) \propto p(y_{t}\vert x_{t},\theta) p(x_{t}\vert \theta) p(\theta),
\end{equation}
which is the basis for full Bayesian estimation procedures. Next we will describe the steps needed for Bayesian DLM estimation of model states, parameters and their uncertainties.

\section{State and parameter estimation}
\label{sec:orgbc28c0d}
To recall the notation, \(y_{t}\) are the observations and \(x_{t}\) are the hidden system states for time indexes \(t=1,\dots,n\). In addition, we have a static vector \(\theta\) that contains auxiliary parameters needed in defining the system matrices \(\mat{M}_{t}\) and \(\mat{H}_{t}\) and the model and observation error covariance matrices \(\mat{Q}_{t}\) and \(\mat{R}_{t}\). For dynamic linear models we have efficient and well founded computational tools for all relevant statistical distributions of interest. For the state estimation assuming a known parameter vector \(\theta\) the assumptions on linearity and Gaussian errors allows us to estimate the model states by classical recursive Kalman formulas. The variance and other structural parameters appear in non-linear way and their estimation can be done either by numerical optimization or by Markov chain Monte Carlo (MCMC) methods. MCMC allows for a full Bayesian statistical analysis for the joint uncertainty in the dynamic model states and the static structural parameters \citep{gamerman06}. Table \ref{tab:org6740377} relates the different statistical distributions to the algorithms, which are outlined later. The notation \(y_{1:t}\), \(x_{1:t}\), etc.\  means the collection of observations or states from time 1 to time \(t\).

\begin{table}[htbp]
\caption{\label{tab:org6740377}
Conditional DLM distributions and the corresponding algorithms. The variables usead are: \(x_{t}\) for the time varying state of the system (e.g.\  trend), \(y_{t}\) for the observations at each time \(t\), and \(\theta\) for structural parameters used in the model and covariance matrices. Notation \(x_{1:n}\) means all time instances for \(1,\dots,n\).}
\centering
\begin{tabular}{lll}
\hline
distribution & meaning & algorithm\\
\hline
\(p(x_{t} \vert x_{t-1},y_{1:t-1},\theta)\) & one step prediction & Kalman filter\\
\(p(x_{t} \vert y_{1:t},\theta)\) & filter solution & Kalman filter\\
\(p(x_t \vert y_{1:n},\theta)\) & smoother solution & Kalman smoother\\
\(p(x_{1:n} \vert y_{1:n},\theta)\) & full state given parameters & simulation smoother\\
\(p(y_{1:t} \vert \theta)\) & marginal likelihood for parameters & Kalman filter likelihood\\
\(p(x_{1:n},\theta \vert y_{1:n})\) & full state and parameter & MCMC\\
\(p(\theta \vert y_{1:n})\) & marginal for parameter & MCMC\\
\(p(x_{1:n} \vert y_{1:n})\) & marginal for full state & MCMC\\
\hline
\end{tabular}
\end{table}

\section{Recursive Kalman formulas}
\label{sec:org24f889f}
Below we give the relevant parts of the recursive formulas for Kalman filter and smoother to estimate the conditional distributions of DLM states given the observations and static parameters. For more details, see \cite{rodgers00,dlmacp}. A notable feature of the linear Gaussian case is that the formulas below are exact and easily implemented in computer as long as the model state dimension or the number of observations at one time is not too large. 

To start the calculations, we assume that the initial distribution of \(x_{0}\) at \(t=0\) is available. The first step in estimating the states is to use Kalman filter forward recursion to calculate the distribution of the state vector \(\state_t\)  given the observations up to time \(t\),
\(p(\state_{t}|\vec{y}_{1:t},\vec{\theta}) = N(\widebar{\state}_{t},\widebar{\mat{C}}_{t})\), which is Gaussian by the linearity assumptions.
At each time \(t\) this step consists of first calculating, as prior, the mean and covariance matrix of one-step-ahead predicted states \(p(\state_{t}|\state_{t-1},\vec{y}_{1:t-1},\vec{\theta}) = N(\widehat{\state}_{t},\widehat{\mat{C}}_{t})\) and the covariance matrix of the predicted observations \(\widehat{\mat{C}}_{y,t}\) as 
\begin{align}
  \widehat{\state}_{t} &= \mat{M}_t\widebar{\state}_{t-1} &\text{prior mean for $x_{t}$,}\label{eq:xpt}\\
  \widehat{\mat{C}}_{t} &= \mat{M}_t \widebar{\mat{C}}_{t-1} \mat{M}_t^{T} + \mat{Q}_t & \text{prior covariance for $x_{t}$,}\\
  \widehat{\mat{C}}_{y,t} &= \mat{H}_t \widehat{\mat{C}}_{t} \mat{H}_t^{T} + \mat{R}_t& \text{covariance for predicting $y_{t}$.}\label{eq:Cyt}
\end{align}
Then the posterior state mean and its covariance are calculated using the Kalman gain matrix \(\mat{G}_t\) as
\begin{align}
  \mat{G}_t &=  \widehat{\mat{C}}_{t} \mat{H}_t^{T}\widehat{\mat{C}}_{y,t}^{-1} &\text{Kalman gain,}\label{equ:gain}\\
  \vec{r}_t &= \vec{y}_{t} - \mat{H}_t\widehat{\state}_{t} & \text{prediction residual,}\\
  \widebar{\state}_{t} &= \widehat{\state}_{t} + \mat{G}_t \vec{r}_t&\text{posterior mean for $x_{t}$,}\label{eq:xhat}\\
  \widebar{\mat{C}}_{t} &= \widehat{\mat{C}}_{t}-\mat{G}_t\mat{H}_t\widehat{\mat{C}}_{t} &\text{posterior covariance for $x_{t}$}.\label{eq:kf2.end}
\end{align}
These equations are iterated for \(t=1,\dots,\nobs\) and the values of \(\widebar{{x}}_{t}\) and \(\widebar{\mat{C}}_{t}\) are stored for further calculations. As initial values, we can use \(\widebar{\state}_{0}=\vec{0}\) and \(\widebar{\mat{C}}_{0}=\kappa \mat{I}\), i.e.\  a vector of zeros and a diagonal matrix with some large value \(\kappa\) in the diagonal. Note that the only matrix inversion required in the above formulas is the one related to the observation prediction covariance matrix \(\widehat{\mat{C}}_{y,t}\), which is of size \(1\times1\) when we analyse univariate time series.

The Kalman filter provides distributions of the states at each time \(t\) given the observations up to the current time. As we want to do retrospective time series analysis that accounts for all of the observations, we need to have the distributions of the states for each time, given all the observations  \(\vec{y}_{1:\nobs}\). By the linearity of the model, these distributions are again Gaussian,
\(p(\state_t|\vec{y}_{1:\nobs},\vec{\theta}) = N(\widetilde{\state}_t,\widetilde{\mat{C}}_t)\).
Using the matrices generated by the Kalman forward recursion,
the Kalman smoother backward recursion gives us the smoothed states for \(t=\nobs,\nobs-1,\dots,1\). There are several equivalent versions of the backward recursion algorithm. Below we show the Rauch-Tung-Striebel recursion \citep{sarkka13} for illustration. For alternatives, see \cite{durbin2012}:

\begin{align}
 \mat{C}_{t}^{+} &= \mat{M}_t \widebar{\mat{C}}_{t} \mat{M}_{t}^{T} + \mat{Q}_{t} & \text{propagated covariance,}\\
    \mat{G}_{t} &= \widebar{\mat{C}}_{t}\mat{M}_{t}^{T} \left(\mat{C}_{t}^{+}\right)^{-1} &\text{smoother gain,}\\
    \widetilde{x}_{t-1} &= \widebar{x}_t + \mat{G}_{t}\left(\widetilde{x}_t - \widebar{x}_t\right) &\text{smoothed state mean,}\\
    \widetilde{\mat{C}}_{t-1} &= \widebar{\mat{C}}_t - \mat{G}_{t}\left( \widetilde{\mat{C}}_t - {\mat{C}}_{t}^{+}\right)\mat{G}_{t}^{T} &\text{smoothed state covariance.}
\end{align}

In smoother recursion we are dealing with several matrix vector computations and one matrix inversion of size \(m\times m\) and these formulas can be implemented quite efficiently in any general numerical analysis software. As a note, we see that the algorithms work with missing observations, too. If some observations at a time \(t\) are missing, the corresponding columns of the gain matrix Eq. (\ref{equ:gain}) will be zero. If all are missing, the filter posterior will be equal to the prior. Note that the above smoother recursion does not refer to the observations. All the Kalman formulas given above are for observations with uniform sampling in time, for non-uniform temporal sampling, the propagation of uncertainty to the next observation time has to be handled differently, see \cite{harvey91,durbin2012}.
\section{Simulation smoother}
\label{sec:org26be235}
The Kalman smoother algorithm provides the distributions \(p(x_{t}|y_{1:n},\theta)\) for each \(t\), which are all Gaussian. However, for studying trends and other dynamic features in the system, we are interested in the joint distribution spanning the whole time range \(p(x_{1:n}|y_{1:n},\theta)\). Note that we are still conditioning on the unknown parameter vector \(\theta\) and will account for it later. This high dimensional joint distribution is not easily accessible directly. As in many cases, instead of analytic expressions, it is more important to be able to draw realizations from the distribution and use the sampling distribution for statistical analysis. This has several benefits. One important is that by comparing simulated realizations to the observations, we see how realistic the model predictions are, which can reveal if the modelling assumptions are not valid. Also, we can study the distributions of model outputs directly from the samples and do not need to resolve to approximate statistics.

A simple simulation algorithm by \cite{durbin2012} is the following. The state space system equations provide a direct way to recursively sample realizations of both the states \(x_{1:n}\) and the observations \(y_{1:n}\), but the generated states will be independent of the original observations. However, it can be shown \citep[Section 4.9]{durbin2012}  that the distribution of the residual process of generated against smoothed state does not depend on \({y}_{1:n}\). This means that if we add simulated residuals over the original smoothed state \(\widetilde{x}_{1:n}\), we get a new realization that is conditional on the original observations \(y_{1:n}\). A procedure to sample a realization \({x}^*_{1:n} \sim p(x_{1:n}|y_{1:n},\theta)\) is thus:

\begin{enumerate}
\item Generate a sample using the state space system equations, Eqs. (\ref{dlmDef1}) and (\ref{dlmDef2}) to get \(\check{x}_{1:n}\) and \(\check{y}_{1:n}\).
\item Smooth \(\check{y}_{1:n}\) to get \({\breve{x}}_{1:n}\) according to formulas in Section \ref{sec:org24f889f}.
\item Add the residuals from step 2 to the original smoothed state:
\end{enumerate}
\begin{equation}
   {x}^*_{1:n} = \check{x}_{1:n} - {\breve{x}}_{1:n} + \widetilde{x}_{1:n}.
\end{equation}

This simulation smoother can be used in trend studies and as a part of more general MCMC simulation algorithm that will sample from the joint posterior distribution \(p(x_{1:n},\theta|y_{1:n})\) and by marginalization argument also from \(p(x_{1:n}|y_{1:n})\) where the uncertainty in \(\theta\) has been integrated out \citep{dlmacp}.

\section{Estimating the static structural parameters}
\label{sec:org8194105}
In the first examples, the variance parameters defining the model error covariance matrix \(\mat{Q}_t\) were assumed to be known. In practice we need some estimation methodology for them. Basically there are three alternatives. The first one uses subject level knowledge with trial and error to fix the parameters without any algorithmic tuning. The second one use the marginal likelihood function with a numerical optimization routine to find the maximum likelihood estimate of the parameter \(\theta\) and plug the estimate back to the equations and re-fit the DLM model. The third one use MCMC algorithm to sample from the posterior distribution of the parameters to estimate the parameters and to integrate out their uncertainty.

To estimate the free parameters \(\theta\) in the model formulation by optimization or by MCMC we need the marginal likelihood function \(p(y_{1:n}|\theta)\). By the assumed Markov properties of the system, this can be obtained sequentially as a byproduct of the Kalman filter recursion \citep{sarkka13},
\begin{equation}\label{eq:kflik}
  -2\log\left(p(y_{1:n}|\theta)\right) = \mathrm{constant}+ \sum_{t=1}^n\left[(y_t-\mat{H}_t\widehat{x}_{t})^T\widehat{\mat{C}}_{y,t}^{-1}(y_t-\mat{H}_t\widehat{x}_{t})
        + \log(|\widehat{\mat{C}}_{y,t}|)\right].
\end{equation}
On the right hand side, the parameter \(\theta\) will appear in the model predictions \(\widehat{x}_{t}\) as they depend on the matrix \(\mat{M}_t\) as well as on the model error \(\mat{Q}_t\). For the same reason we need the determinant of the model prediction covariance matrix \(|\widehat{\mat{C}}_{y,t}|\). A fortunate property is that this likelihood can be calculated along the DLM filter recursion without much extra effort. 

The scaled one-step prediction residuals 
\begin{equation}\label{eq:resid}
	r^{*}_{t} = \widehat{\mat{C}}_{y,t}^{-1/2}(y_t-\mat{H}_t\widehat{x}_{t})
\end{equation}
can be used to check the goodness of fit of the model. In order of the DLM model to be consistent with the observations these residuals should be approximately independent, \(N(0,\mat{I})\) Gaussian and without serial autocorrelation. Later in the GNSS time series example, we will do model diagnostics by residual quantile-quantile and autocorrelation function plots.

\section{Analysing trends}
\label{sec:orgcf0c233}
In general terms, trend is a change in the distributional properties, such as in the mean, of the process generating the observations. We are typically interested in slowly varying changes in the background level, i.e.\  in the mean process after the known sources of variability, such as seasonality, has been accounted for. A common way to explore trends is to fit some kind of a smoother, such as a moving average, over the time series. However, many standard smoothing methods do not provide statistical estimates of the smoothness parameters or asses the uncertainty related to the level of smoothing.

In typical DLM trend analyses, a slowly varying (relative to the time scale we are interested in) background level of the system is modelled as a first or higher order random walk process with variance parameters that determine the time wise smoothness of the level. These variance parameters must be estimated and their uncertainty accounted for proper uncertainty quantification. In an optimal case, the data provides information on the smoothness of the trend component, but typically we need to use subject level prior information to decide the time scale of the changes we want to extract. 
In the GNSS application example in Section \ref{sec:orga943c82} we assume a global linear trend and the local non-stationary fluctuations are modelled using a local random walk model with autocorrelated residuals.
A Bayesian DLM model offers means to provide qualitative prior information in the form of the model equations and quantitative information by prior distributions on the variance parameters, see e.g.\  \citet{gamerman06}.

For statistical analysis we need to estimate the full state as either \(p(x_{1:n}|y_{1:n},\hat{\theta})\), where we plug in some estimates of the auxiliary parameters \(\hat{\theta}\), (the maximum likelihood approach) or by \(p(x_{1:n}|y_{1:n}) = \int p(x_{1:n},\theta|y_{1:n})\,d\theta\) where the uncertainty of auxiliary parameter \(\theta\) is integrated out. The latter is  the Bayesian approach and calculations can be done, e.g., by Markov chain Monte Carlo (MCMC) simulation \citep{gamerman06,dlmacp}. 
A procedure to account the uncertainty in a DLM model and its structural parameters and to study DLM output will contain the following steps:

\begin{enumerate}
\item Formulate the DLM model and its marginal likelihood \(p(y_{1:n} \vert \theta)\) by Kalman filter.
\item Use MCMC to sample from the posterior distribution \(p(\theta\vert y_{1:n})\) with a suitable prior distribution \(p(\theta)\) for the structural parameters and with the likelihood of step 1.
\item Generate a sample from the marginal posterior \(p(x_{1:n}\vert y_{1:n})\) using the simulation smoother (Section \ref{sec:org26be235}) and for each sample use a different \(\theta\) from the MCMC chain from the previous step.
\item For each state realization, \(x^{*}_{1:n} \sim p(x_{1:n}\vert y_{1:n})\), from step 3., calculate a trend related or any other statistics of interest and use this sample for the estimates and their uncertainties.
\end{enumerate}

\section{Examples of different DLM models}
\label{sec:org5812479}
In the following, we give several useful DLM formulations for model components that are typically used in geodetic or in more general environmental analyses. They have been used in existing applications for stratospheric ozone \citep{dlmacp}, ionosonde analysis \citep{ionosonde} and for station temperature records \citep{mikkonen14}. In Section \ref{sec:orga943c82}, we will show analysis for synthetic GNSS station positioning time series.

\subsection{The effect of level and trend variance parameters}
\label{sec:org16e5353}
\begin{figure}[htbp]
\centering
\includegraphics[width=.9\linewidth]{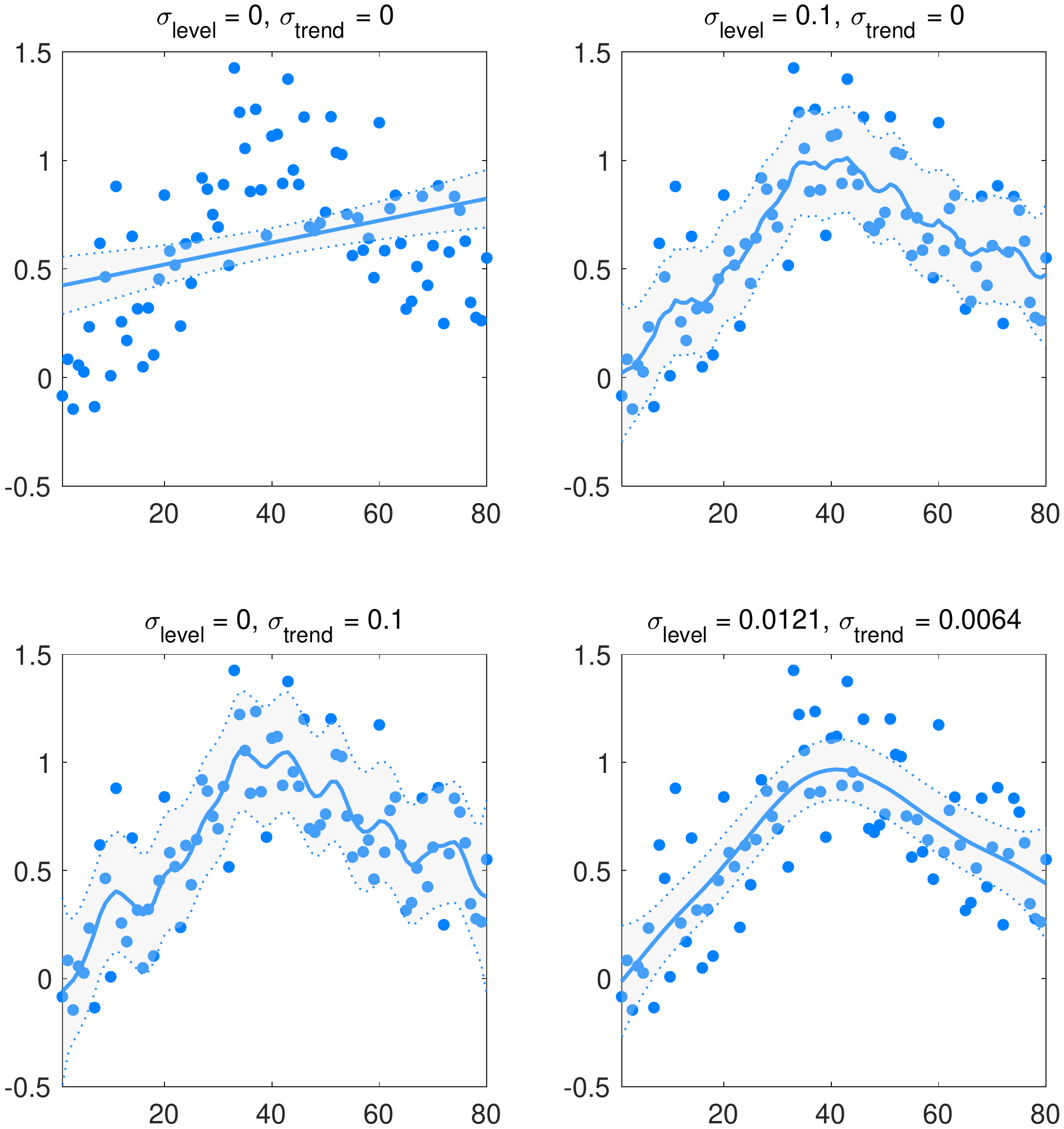}
\caption{\label{fig:orge557686}
DLM smoother fit for sythetic data set with different smoothing levels. The dots are the observations and solid blue line is the mean DLM fit. The grey area corresponds to 95\% probability limit from the Kalman smoother. The last panel uses the parameter obtained by MCMC.}
\end{figure}

In the first example in Section \ref{sec:org0f86055} the variance \(\sigma^2_\mathrm{trend}\) was assumed to be known and fixed. Altering the variance affects the smoothness of the fit. In Figure \ref{fig:orge557686} the effect of different variance parameters are shown for the same data. Note that by setting both \(\sigma^2_\mathrm{level}\) and \(\sigma^2_\mathrm{trend}\) to zero results in classical linear regression without dynamical evolution of the regression components. In this case, the 95\% probability limits for the level obtained from the smoother covariance matrix \(\widetilde{\mat{C}}_{t}\) coincide with the classical confidence intervals for the mean. In classical non-dynamic linear regression the modelling error is included in the residual term, whereas in DLM we can include it in the model definition by allowing temporal change in model parameters.

\begin{figure}[htbp]
\centering
\includegraphics[width=.9\linewidth]{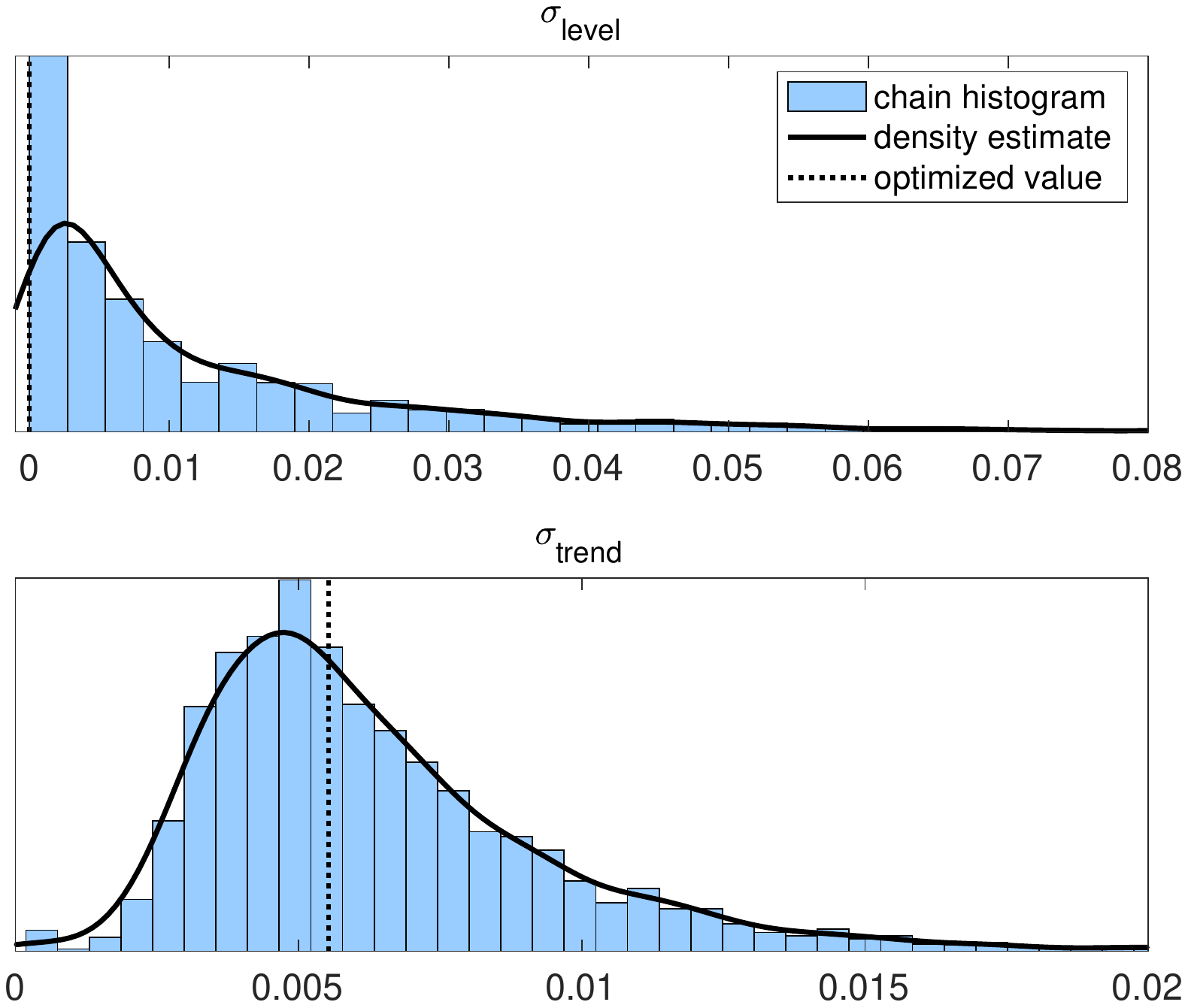}
\caption{\label{fig:org02e6d84}
Two variance parameters of example \ref{sec:org16e5353} estimated by MCMC. Histogram is the MCMC chain histogram. The solid line is a kernel estimate for the marginal posterioir probability distribution. Dotted vertical line is obtained by numerical minimization of the log likelihood.}
\end{figure}

If we estimate the parameters by the likelihood approach and MCMC outlined in Section \ref{sec:org8194105}, we get the values in the last panel of Figure \ref{fig:orge557686} corresponding to the posterior mean. Figure \ref{fig:org02e6d84} shows MCMC chain histograms together with estimated marginal posterior densities. It also has the point values obtained by likelihood optimization. Note by optimization we get an estimate for \(\sigma_\mathrm{level}\) which is very close to zero relatively to the MCMC solution, which tries to find all values of the parameter that are consistent with the data.

\subsection{Seasonal component}
\label{sec:org37a717a}
Seasonal variability can be modelled by adding extra state components for the effect of each season. A common description of seasonality uses trigonometric functions and is achieved by using two model states for each harmonic component. Monthly data with annual and semiannual cycles would use four state components and the following model and observation matrices

\begin{equation}\label{eq:sG}
\mat{M}_{\mathrm{seas}}=\begin{bmatrix}
 \cos(\pi/6) & \sin(\pi/6) & 0 & 0\\
-\sin(\pi/6) & \cos(\pi/6) & 0 & 0\\
0 & 0 &             \cos(\pi/3) & \sin(\pi/3) \\
0 & 0 &             -\sin(\pi/3) & \cos(\pi/3) \\
\end{bmatrix}
\end{equation}
and
\begin{equation}\label{eq:sF}
\mat{H}_{\mathrm{seas}}= \begin{bmatrix}1 & 0 & 1& 0\\\end{bmatrix}.
\end{equation}

In addition, a corresponding part or the model error covariance matrix \(\mat{Q}_\mathrm{seas}\) has to be set up to define the allowed variability in the seasonal amplitudes. A simple approach is to use a diagonal matrix with equal values for each component as \(\mathrm{diag}(\mat{Q}_\mathrm{seas}) = [\sigma^2_\mathrm{seas},\sigma^2_\mathrm{seas},\sigma^2_\mathrm{seas},\sigma^2_\mathrm{seas}]^{T}\). If we set these variances to zero, the DLM algorithm will fit a temporally fixed seasonal amplitude.

\begin{figure}[htbp]
\centering
\includegraphics[width=.9\linewidth]{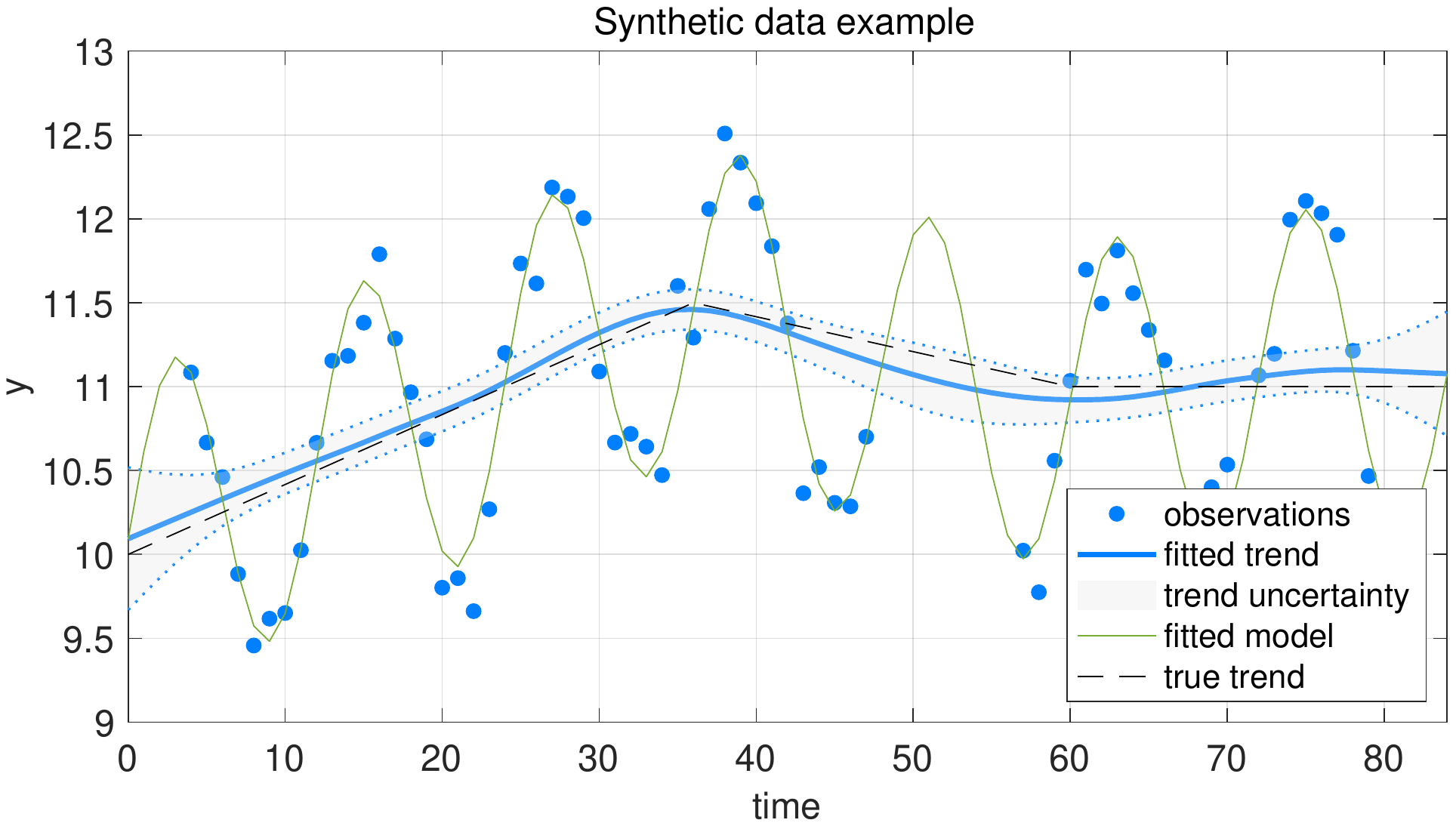}
\caption{\label{fig:org3ad07ad}
DLM smoother fit to synthetic data in Section \ref{sec:org37a717a} with seasonal variation, piecewise linear trend, and missing observations.}
\end{figure}

For illustration we use a simulated monthly data with yearly variation that has some randomness in the amplitude. The observations have a piecewise linear trend similar to example in Section \ref{sec:org0f86055} and some values as missing to see the effect on the uncertainties. We fit a seasonal component with one harmonic function, but we allow some variability in the amplitude and trend, with \(\sigma_\mathrm{trend} = 0.005\) and \(\sigma_\mathrm{seas} = 0.4\). Figure \ref{fig:org3ad07ad} shows the generated data together with both the fitted mean process and the fitted seasonal component. A similar example was also used in \cite{ionosonde}.

\subsection{Autoregressive process}
\label{sec:orgb2b305d}
Autoregressive processes have serial dependence between the observations. A general AR(\(p\)) process is defined by \(p\) coefficients \([\rho_{1},\dots,\rho_{p}]\) and an independent innovation term \(\epsilon\) as
\begin{equation}
y_{t} = \rho_{1}y_{t-1} + \rho_{2}y_{t-2} + \dots + \rho_{p}y_{t-p} + \epsilon, \qquad
\epsilon \sim N(0,\sigma^{2}_\mathrm{AR})
\end{equation}

For including an autoregressive component into the state space formulation we need to use state variables that "remember" their previous values. This can be achieved by suitable evolution operator \(\mat{M}_\mathrm{AR}\). For example, AR(3) process with coefficients \([\rho_{1},\rho_{2},\rho_{3}]\), will have three extra states with
\begin{equation}
\mat{M}_\mathrm{AR} =
  \begin{bmatrix}
     \rho_1 & 1  & 0 \\
     \rho_2 & 0  & 1 \\
     \rho_3 & 0  & 0 \\
  \end{bmatrix},\quad
\mat{H}_\mathrm{AR} =
  \begin{bmatrix}
     1 & 0 & 0 \\
  \end{bmatrix},\quad
\mat{Q}_\mathrm{AR} =
  \begin{bmatrix}
     \sigma^{2}_\mathrm{AR} & 0 & 0 \\
     0 & 0 & 0 \\
     0 & 0 & 0 
  \end{bmatrix}.
\end{equation}

A pure AR(3) process would then be obtained by setting the observation error \(\sigma^{2}_\mathrm{obs}\) in Eq. (\ref{dlmDef1}) to zero and the model error component equal to the innovation variance \(\sigma^{2}_\mathrm{AR}\). 
If we, in addition, have \(\sigma^{2}_\mathrm{obs}>0\), it will result to an ARMA. In fact all ARMA and ARIMA models can be represented as DLM models \citep[Section 3.2.5]{petris2009}  and many ARIMA estimation software implementations use the Kalman filter likelihood Eq. (\ref{eq:kflik}) to formulate the cost function for estimation.

\subsection{Regression covariates and proxy variables}
\label{sec:org4646faa}
In many applications the variability in the observations is affected by some known external factors, such as temperature, air pressure or solar activity. Sometimes these variables can be measured directly, as for the temperature, and sometimes their effect is modelled via a proxy, such as a radio fluxes for the solar effect. As an example, assume an observations model 
\begin{equation}\label{equ:proxy}
y_{t} = \mu_{t} + \gamma_{t} + \beta_{t}\mat{Z}_{t} + \epsilon_\mathrm{obs},
\end{equation}
where \(\mu_{t}\) and \(\gamma_{t}\) are the mean level and the seasonal components, \(\mat{Z}_{t}\) is a row matrix of the values of the regression variables at time \(t\), and \(\beta_{t}\) is a vector of time-varying regression coefficients.
The effect of the covariates can be formulated by having the coefficients as extra states, \(x_{\mathrm{proxy},t} = \beta_{t}\), using an identity model operator, and by adding the covariate values to the observation operator \(\mat{H}_{t}\) as
\begin{align}
 \mat{H}_{\mathrm{proxy}(t)}&= \mat{Z}_{t} = \begin{bmatrix}Z_{t,1},\dots,Z_{t,p}\\\end{bmatrix},\\
 \mat{M}_{\mathrm{proxy}}&= \mat{I}_{p}= \mathrm{diag}(1,\dots,1),\\
 \mat{Q}_{\mathrm{proxy}}&= \mathrm{diag}\left(\begin{bmatrix}\sigma^2_{\mathrm{proxy},1},\dots,\sigma^2_{\mathrm{proxy},2}\\\end{bmatrix}\right).
\end{align}
The DLM model for equation Eq. (\ref{equ:proxy}) is then build up as diagonal block matrix combination of the components:
\begin{align}
x_{t} &= \begin{bmatrix} x_{\mathrm{trend},t}& x_{\mathrm{seas},t}& x_{\mathrm{proxy},t} \end{bmatrix}^{T},\\
    \mat{M}_{t} &=\begin{bmatrix}
      \mat{M}_\mathrm{trend} & 0 & 0 \\
      0 & \mat{M}_{\mathrm{seas}} & 0 \\
      0 &  0 & \mat{M}_\mathrm{proxy}
    \end{bmatrix},\\
    \mat{H}_t &= \begin{bmatrix}\mat{H}_\mathrm{trend}&\mat{H}_{\mathrm{seas}}&\mat{H}_{\mathrm{proxy}(t)}\\\end{bmatrix},\\
    \mat{Q}_{t} &=\begin{bmatrix}
      \mat{Q}_\mathrm{trend} & 0 & 0 \\
      0 & \mat{Q}_{\mathrm{seas}} & 0 \\
      0 &  0 & \mat{Q}_\mathrm{proxy}
    \end{bmatrix}.
\end{align}

The covariate variances \(\sigma^2_{\mathrm{proxy}}\) control the allowed temporal variability in the coefficients \(\beta_{t}\) and their values can be estimated or set to some prior value. By setting the variances to zero, turns this model into classical multiple linear regression. 

\section{Synthetic GNSS example}
\label{sec:orga943c82}
\begin{figure}
  \begin{center}
  \includegraphics[width=0.49\textwidth]{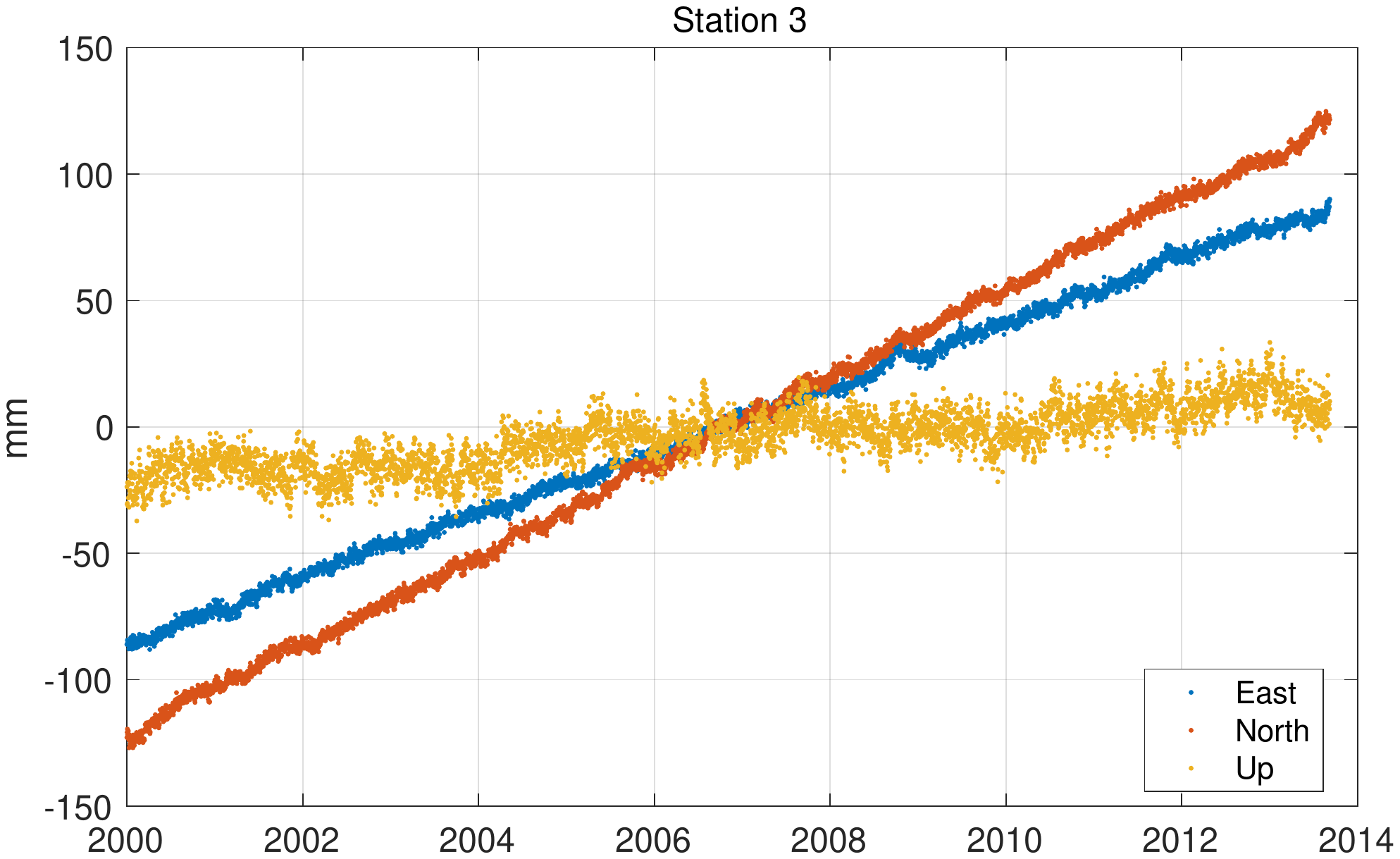}
  \includegraphics[width=0.49\textwidth]{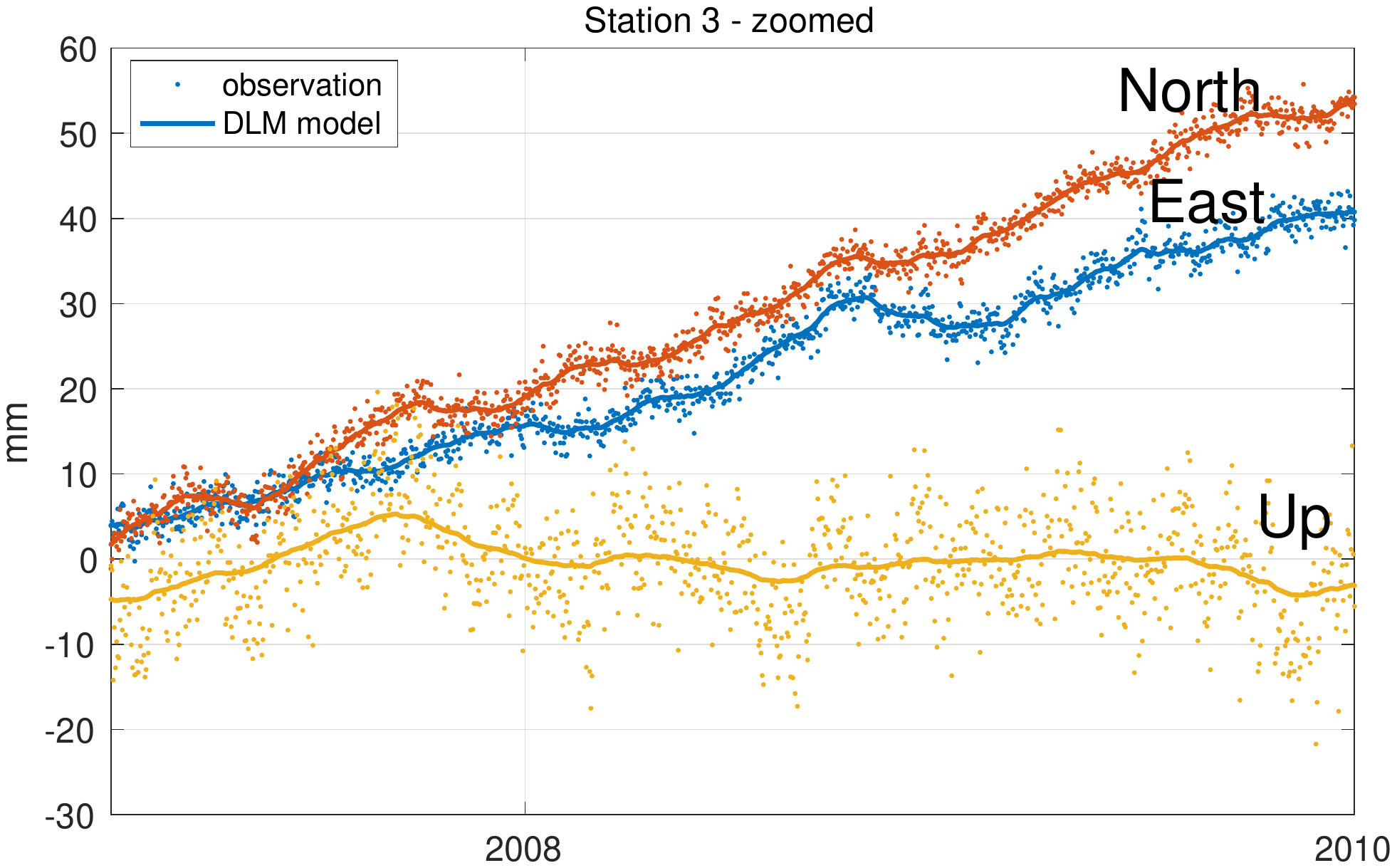}\\
  \includegraphics[width=0.49\textwidth]{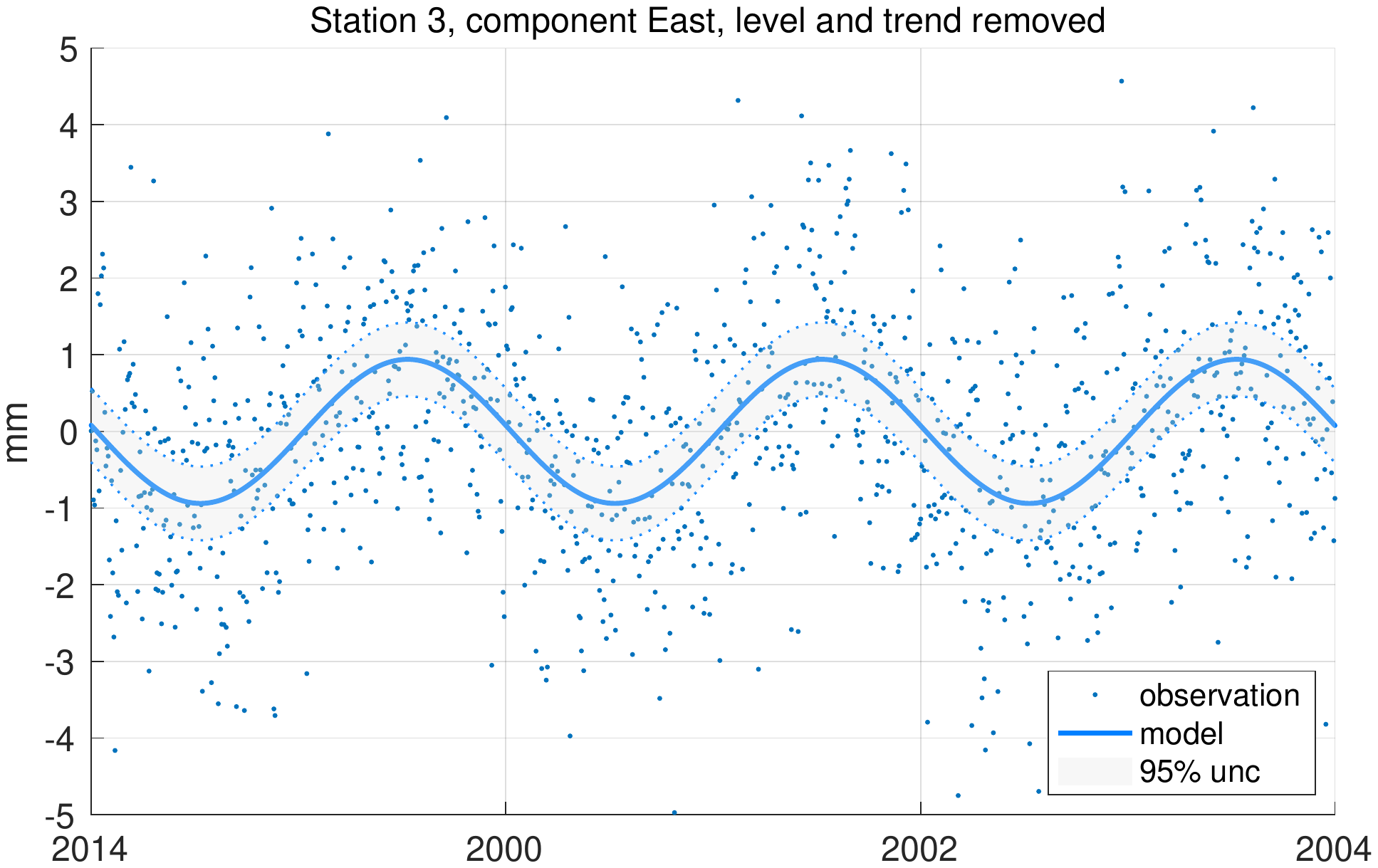}
  \includegraphics[width=0.49\textwidth]{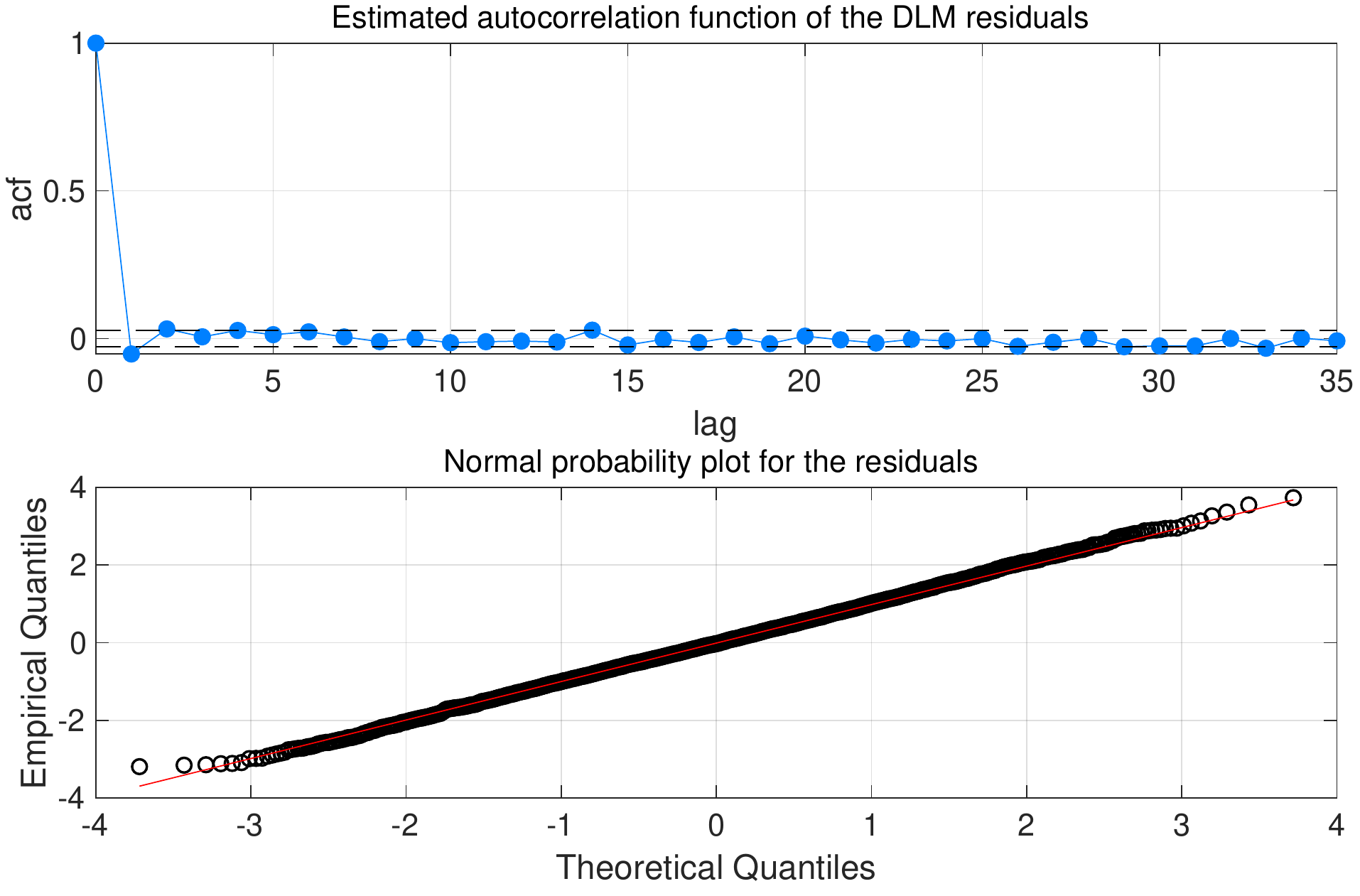}    
  \end{center}
  \caption{GNSS example data set and the DLM fit. Top left: three data components. Top right: zoomed component with DLM fit. Bottom left: The "East" component with the modelled level and trend removed, showing the seasonal variation and the model residual over it. Bottom right: residual diagnostics of the DLM fit for "East".}
  \label{fig:gnss4}
\end{figure}

Next we estimate trends in synthetic GNSS time series provided by Machiel Bos and Jean-Philippe Montillet. In this application, the trend estimated in the GNSS time series represents the tectonic rate on the East and North components and the vertical land motion on the Up coordinate. The characteristics of the GNSS time series are discussed in details in Chapter 1 and 2. We select data for one of the stations (labeled station n:o 3 in the figures) with the three components (East, North, Up) shown in Figure \ref{fig:gnss4}, top left panel. The time series are simulated using linear trend, yearly seasonal variation and a combination of coloured and i.i.d Gaussian noise. We assume that we do not know the noise structure a priori. We are interested in the (non-local) linear trend and we need a model component for the local fluctuations seen in the data.  This chosen data sets does not contain any sudden jumps in the measured position. Modelling offset changes would require a different strategy, with some iterative estimate of the jump locations, which we will not consider here.
We use a DLM approach, where we assume that the non-stationary part can be modelled by local polynomials and the stochastic stationary part can be described as an AR or ARMA process in addition to the i.i.d.\  Gaussian observation uncertainty.
See \cite{Dmitrieva2015} for a somewhat similar approach, which uses state space representation and Kalman filter likelihood to model flicker and random walk type noise in several stations at the same time.

So, in contrast to the spline smoothing example in Section \ref{sec:org0f86055}, which had \(\sigma^{2}_\mathrm{level} = 0\) and \(\sigma^{2}_\mathrm{trend} > 0\), we will extract a non-local linear trend, \(\sigma^{2}_\mathrm{trend} = 0\), and model the local non-stationary fluctuations as a local level model with \(\sigma^{2}_\mathrm{level} > 0\). In addition, we use a yearly seasonal component for the daily observations and an autoregressive AR(1) noise component to account for the possible residual correlation. The observation error is assumed Gaussian and to have known standard deviation, \(\sigma_\mathrm{obs} = 1 \mathrm{mm}\) for components "East" and "North" and \(\sigma_\mathrm{obs} = 4 \mathrm{mm}\) for the "Up" component. The AR(1) innovation variance \(\sigma_\mathrm{AR}\) as well as the AR coefficient \(\rho_\mathrm{AR}\) will be estimated from the data. We use Kalman filter likelihood to estimate the 2 variance parameters and the AR(1) coefficient by MCMC. We analyse the three components (East, North, Up) separately. 

\begin{figure}
  \begin{center}
  \includegraphics[width=0.49\textwidth]{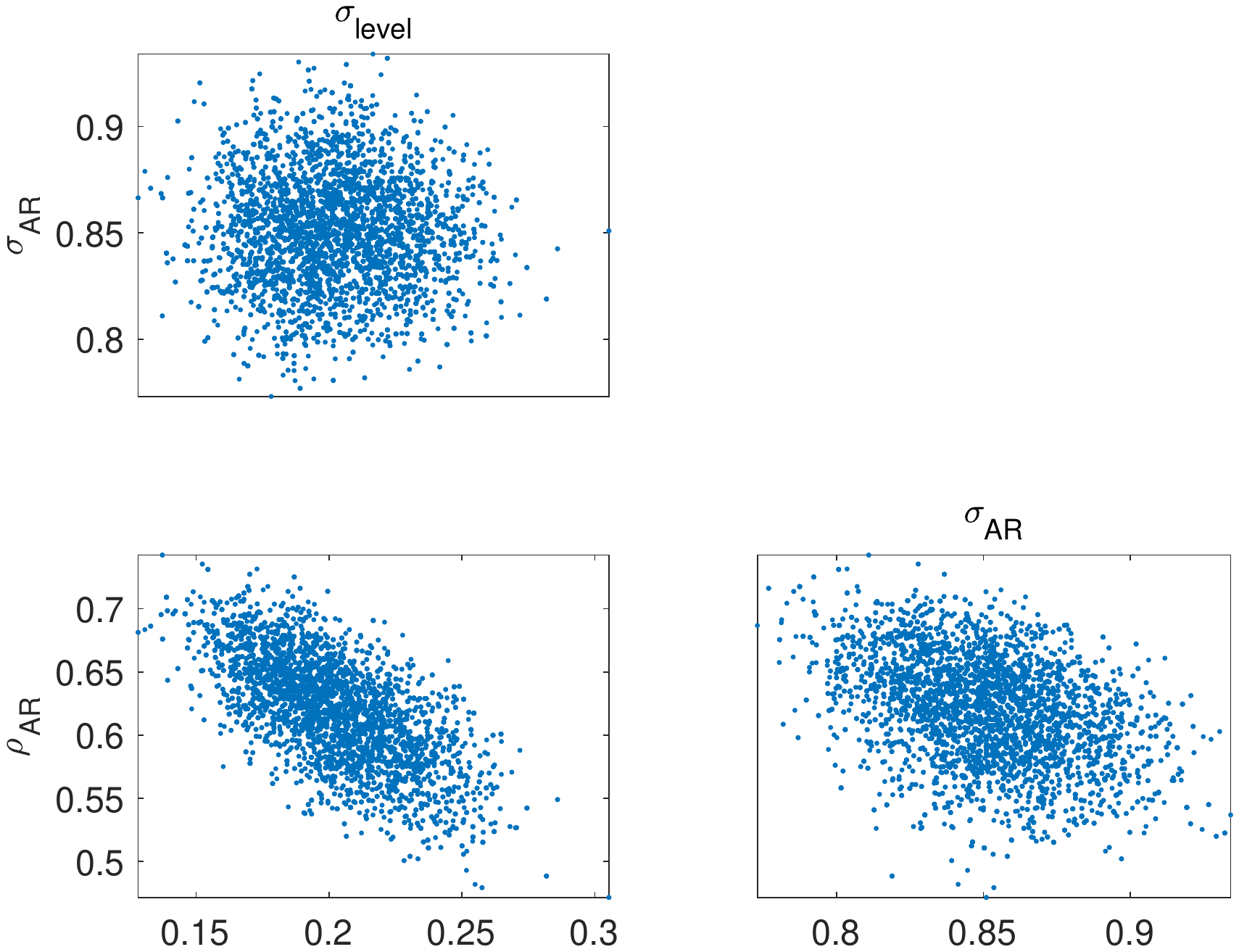}
  \includegraphics[width=0.49\textwidth]{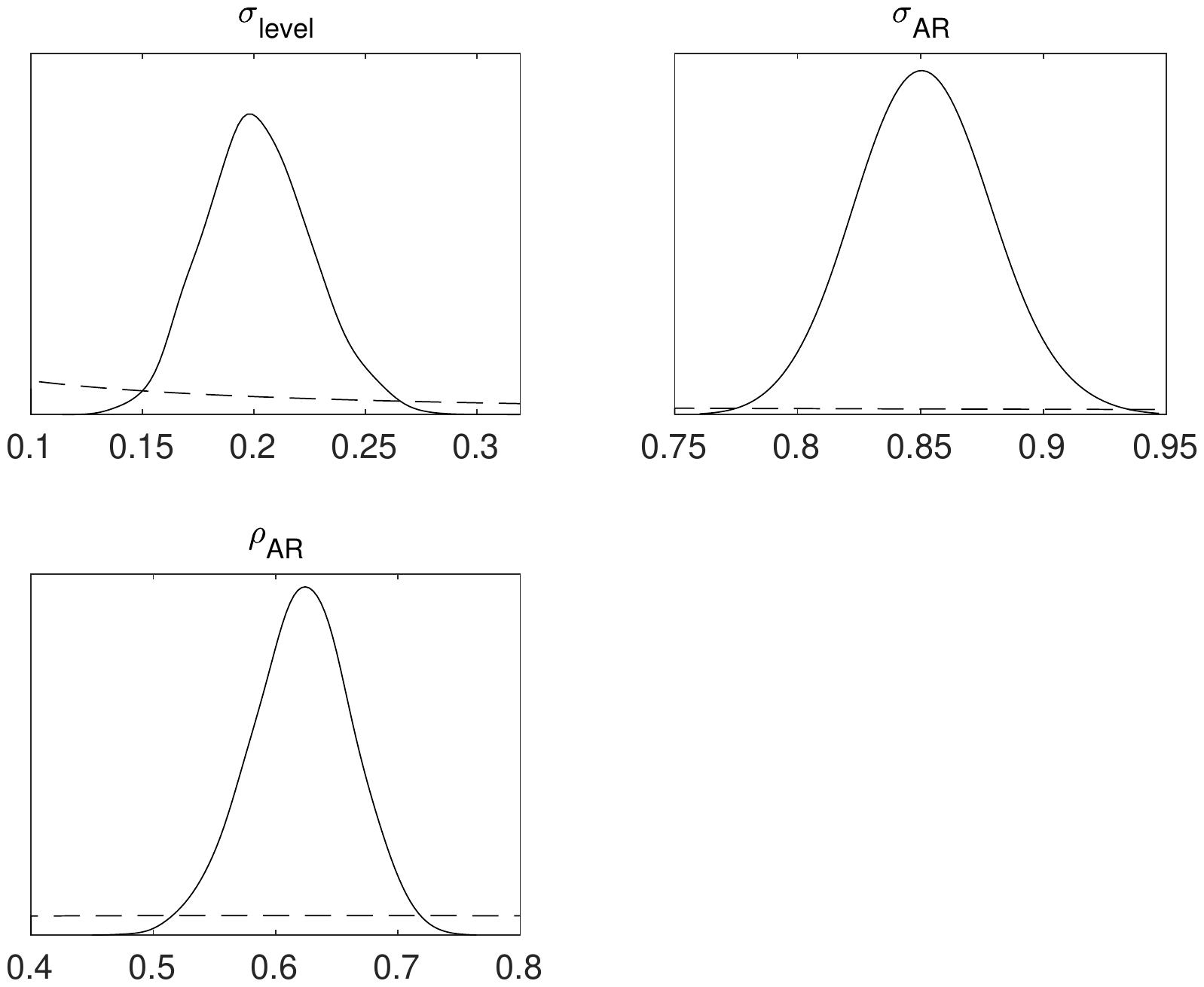}    
  \end{center}
  \caption{GNSS example data set and the DLM fit. On left are the pairwise scatter plots of the MCMC samples for the model parameters for the "East" observations. Right panel shows the estimated marginal posterior densities. The dashed line is the corresponding prior density used.}
  \label{fig:gnssmcmc}
\end{figure}

The true trend coefficients used in the simulation for the three data sets were give as 12.59, 17.64, and 2.778 mm/yr. The estimates obtained for them were \(12.62 \pm 0.61\), \(17.76 \pm 0.69\) and \(2.22 \pm 1.00\) mm/yr, with one-sigma posterior standard deviations after \(\pm\).
Table \ref{tab:org795d1be} shows the parameter estimates obtained by combination of Kalman simulation smoother for the linear slope and seasonal amplitude, and MCMC for \(\theta = [\sigma_\mathrm{level}, \sigma_\mathrm{AR}, \rho_\mathrm{AR}]^{T}\). Figures \ref{fig:gnss4} and \ref{fig:gnssmcmc} visualise the results graphically. There is a hint of negative autocorrelation in the ACF plot for the East components in Figure \ref{fig:gnss4}, but otherwise the residuals, obtained from the scaled prediction residuals, equation Eq. (\ref{eq:resid}), look very Gaussian. In overall, the selected DLM model seems to provide statistically consistent fit and reproduce the true trends within the estimated uncertainty.

\begin{table}[htbp]
\caption{\label{tab:org795d1be}
Parameter estimates from DLM/MCMC estimation for the synthetic GNSS time series example. The uncertainty value is one-sigma posterior standard deviation. The true values for trends were 12.59, 17.64, and 2.778 mm/yr. The true seasonal amplitude was 1 mm.}
\centering
\begin{tabular}{llllll}
data & trend [mm/yr] & seasonal [mm] & \(\sigma_\mathrm{level}\) & \(\sigma_\mathrm{AR}\) & \(\rho_\mathrm{AR}\)\\
\hline
East & \(12.62\pm0.61\) & \(0.93\pm0.15\) & \(0.20\pm0.024\) & \(0.85\pm0.024\) & \(0.62\pm0.03\)\\
North & \(17.76\pm0.69\) & \(1.19\pm0.16\) & \(0.22\pm0.02\) & \(0.86\pm0.024\) & \(0.64\pm0.29\)\\
Up & \(2.22\pm1.00\) & \(0.74\pm0.29\) & \(0.34\pm0.07\) & \(2.00\pm0.075\) & \(0.87\pm0.016\)\\
\end{tabular}
\end{table}

\section{Computer implementation}
\label{sec:org35dd06f}

The examples and code to fit DLM models described here are available from a Github repository at \url{https://github.com/mjlaine/dlm}. The code is written in Matlab and it contains a reference implementation of Kalman filter, smoother and simulator algorithms as well as optimization and MCMC for the structural parameters. Other software implementations for DLM include state space models toolbox for Matlab described in \cite{JSSv041i06}, a R package \texttt{dlm} described in \cite{petris2009} and python implementations in the \texttt{statsmodes} package \citep{statsmodels}.

\section{Conclusions}
\label{sec:orgc73948d}

DLM provides a general framework for modelling many kinds of environmental time series, including geodetic ones. Some features of GNSS time series, such as the often assumed flicker noise and handling of offsets and data jumps might still require more special treatments. However, the DLM approach provides a very useful  generalization to the ordinary linear regression model. Its strengths include the ability to model non-stationary processes by allowing temporal change in the model coefficients and the direct modelling of the processes that generate the observed variability. By guiding the analysis in terms of the generating processes and their uncertainties it provides a good basis for Bayesian statistical inference. If there is prior knowledge about the changes, such as known change points in the data, they can be included in the model. By using simulation based Bayesian DLM analysis, your prior and posterior model simulations can be checked to be consistent with physical constrains and the observations.

\section*{References}
\label{sec:org76f575f}
\renewcommand{\bibsection}{}
\bibliographystyle{spbasic}
\bibliography{master}
\end{document}